\documentclass{article}

\usepackage{arxiv}
\usepackage[english]{babel}
\usepackage{natbib}

\usepackage[utf8]{inputenc} 
\usepackage[T1]{fontenc}    
\usepackage{hyperref}       
\usepackage{url}            
\usepackage{booktabs}       
\usepackage{amsfonts}       
\usepackage{nicefrac}       
\usepackage{microtype}      
\usepackage{lipsum}
\usepackage{graphicx}
\usepackage{float}

\title{Landslide Segmentation with U-Net: Evaluating Different Sampling Methods and Patch Sizes}

\author{
  Lucas P. Soares \thanks{Corresponding Author \newline  \textsuperscript{a} Spatial Analysis and Modelling Lab (SPAMLab, IEE-USP) - \url{https://spamlab.github.io} }\space \textsuperscript{a} \\
  Institute of Geosciences  \\
  University of São Paulo (USP)\\
  São Paulo, Brazil \\
  \texttt{lpsoares@usp.br} \\
  \And
 Helen C.~Dias \textsuperscript{a} \\
  Institute of Energy and Environment\\
  University of São Paulo (IEE-USP)\\
  São Paulo, 05508-010, Brazil \\
  \texttt{helen.dias@usp.br} \\
     \And
 Carlos H.~Grohmann\textsuperscript{a} \\
  Institute of Energy and Environment\\
  University of São Paulo (IEE-USP)\\
  São Paulo, 05508-010, Brazil \\
  \texttt{guano@usp.br} \\
}

\begin{document}
\maketitle

\begin{abstract}
Landslide inventory maps are crucial to validate predictive landslide models; however, since most mapping methods rely on visual interpretation or expert knowledge, detailed inventory maps are still lacking. This study used a fully convolutional deep learning model named U-net to automatically segment landslides in the city of Nova Friburgo, located in the mountainous range of Rio de Janeiro, southeastern Brazil. The objective was to evaluate the impact of patch sizes, sampling methods, and datasets on the overall accuracy of the models. The training data used the optical information from RapidEye satellite, and a digital elevation model (DEM) derived from the L-band sensor of the ALOS satellite. The data was sampled using random and regular grid methods and patched in three sizes (32x32, 64x64, and 128x128 pixels). The models were evaluated on two areas with precision, recall, f1-score, and mean intersect over union (mIoU) metrics. The results show that the models trained with 32x32 tiles tend to have higher recall values due to higher true positive rates; however, they misclassify more background areas as landslides (false positives). Models trained with 128x128 tiles usually achieve higher precision values because they make less false positive errors. In both test areas, DEM and augmentation increased the accuracy of the models. Random sampling helped in model generalization. Models trained with 128x128 random tiles from the data that used the RapidEye image, DEM information, and augmentation achieved the highest f1-score, 0.55 in test area one, and 0.58 in test area two. The results achieved in this study are comparable to other fully convolutional models found in the literature, increasing the knowledge in the area.
\end{abstract}


\section{Introduction}
Natural hazards are more frequent and harmful in recent years due to unplanned urbanization, climate change, and population growth \citep{kobiyama2006prevenccao, hong2017rainfall,alexander2008brief,zhong2019landslide}. According to the Sendai framework for disaster risk reduction 2015-2030 \citep{unisdr2015sendai}, between 2008 and 2012, those hazards affected more than 25 million people, with an economic loss of about 1.3 trillion dollars, impeding the progress towards sustainable development.

 Landslides commonly cause victims, damages to human habitations, and economic losses. Therefore, the study of landslide detection has been considered a critical area of research in remote sensing \citep{hong2017rainfall}. However, despite the importance highlighted by many authors, detailed landslides inventories are still lacking \citep{mondini2019sentinel, guzzetti2012landslide}. Landslide inventory maps are used to prepare and validate landslide susceptibility models, evaluate risk and vulnerability, study erosion and geomorphology, and document the impact of a landslide disaster \citep{VanWesten2008}. Limited and incomplete data may be a source of bias for these studies since model success depends directly on inventory accuracy. 

Landslides inventory maps usually are prepared by using remote sensing imagery with high (HR) and very-high (VHR) resolution \citep{zhong2020landslide}. The landslides can be recognized in an aerial image manually by visual interpretation, semi-automatically, or automatically by using algorithms for object image analysis (OBIA) and pixel-based classification. Manual classification of landslides is the prevailing method \citep{xu2015preparation, yu2020landslide}, but, for large areas, it is time-consuming. OBIA methodologies classify landslide areas by grouping objects with similar spectral, spatial, hierarchical, textural, and morphological properties \citep{Blaschke2010}. Nevertheless, the assignment of those parameters is highly dependent on the analyst experience. Pixel-based methodologies classify each pixel of the image based on its spectral information. However, geometric and contextual information present in the image is ignored, increasing the salt-and-pepper noise in the results \citep{stumpf2011object,blaschke2014geographic,zhong2019landslide,prakash2020mapping}.

In recent years, deep convolution neural networks (DCNN) achieve state-of-art results in applications such as semantic segmentation, object detection, natural language processing, and speech recognition \citep{ghorbanzadeh2019evaluation, peng2019end,zhu2017deep,long2015fully,radovic2017object}. However, only a few studies have used DCNNs for landslide detection \citep{zhong2020landslide}.

\citet{ding2016automatic} used DCNN on GF-1 (Gaofen-1) images with four spectral bands and eight-meter resolution, achieving an overall accuracy of 67\%, a detection rate of 72.5\%, and 10.2\% of false positive rate. \citet{chen2018automated} used DCNN on bi-temporal images to evaluate areas with drastic changes and combined a spatial context learning (STCL) and information from a digital elevation model (DEM)  to detect landslide areas. The method yield an accuracy of more than 61\% on the evaluated areas. \citet{ghorbanzadeh2019evaluation} compared state-of-art machine learning methods and DCNN on RapidEye images and a DEM, with five meters of spatial resolution. The DCNN that used only spectral information and small windows was the best model achieving 78.26\% on the mean intersect over union (mIoU) metric. \citet{sameen2019landslide} compared residual networks (ResNets) trained with topographical information fused using convolutional networks with topographical data added as additional channels. The models trained with the fused data achieved f1-score and mIoU that were superior by 13\% and 12.96\% compared to the other models. \citet{yu2020landslide} used the enhanced vegetation index (EVI), DEM degradation indexes, and a contouring algorithm on Landsat images to sample potential landslide zones with less class imbalance distribution. The trained fully convolutional network (PSPNet) achieved 65\% of recall and 55.35\% of precision.  \citet{prakash2020mapping} used Lidar DEM and Sentinel-2 images to compare traditional pixel, object, and DCNN methods. The deep learning method, U-net with ResNet34 blocks, achieved the best results with the Matthews correlation coefficient score of 0.495 and the probability of detection rate of 0.72.

DCNNs, in supervised learning problems, can learn to identify patterns on the training data without the need for complex operations to extract features or preprocessing methods. However, choosing the best network architecture, preparing the training dataset, and tuning the hyperparameters is still a challenge \citep{pradhan2017integration,sameen2019landslide}. Landslides scars dataset usually have an imbalanced class distribution with more pixels belonging to background objects, such as urban areas, vegetation, and water, than landslide scars \citep{yu2020landslide}. Therefore, since landslide scars have different shapes and sizes, sampling methods and patch sizes may affect the model accuracy as it can be a way to reduce the class imbalance between the positive and the negative class. 

This research aims to evaluate how different datasets, sampling methods, and patch sizes impacts on the landslide segmentation accuracy of U-net. To achieve that, we trained 288 models with landslide optical information from a RapidEye satellite and topographical information from a DEM derived from the Phased Array type L-band Synthetic Aperture Radar (PALSAR) sensor of the ALOS satellite. The models were trained with images patched in three different sizes (32x32, 64x64, 128x128 pixels), and sampled using random and regular grid sampling methods. Data augmentation was also tested. The study area is in the city of Nova Friburgo, located in the mountainous range of Rio de Janeiro, Brazil. The models were evaluated in two test areas with f1-score, recall, precision, and mean intersect over union (mIoU) metrics.

The main contributions of this research are as follows:

\begin{itemize}
\item Broad comparison between patch sizes, sampling method, and datasets.
\item Evaluation of U-net architecture for semantic segmentation of landslides.
\end{itemize}

\section{Study Area}

In January 2011, an extreme rainfall event (350~mm/48h) triggered at least 3500 translational landslides that killed more than 1500 people and disrupted all major city facilities in the mountainous region of Rio de Janeiro, Brazil. This event is considered the worst Brazilian natural disaster \citep{avelar2013mechanisms}.

\begin{figure}[!ht]
\centering
\includegraphics[width=\textwidth]{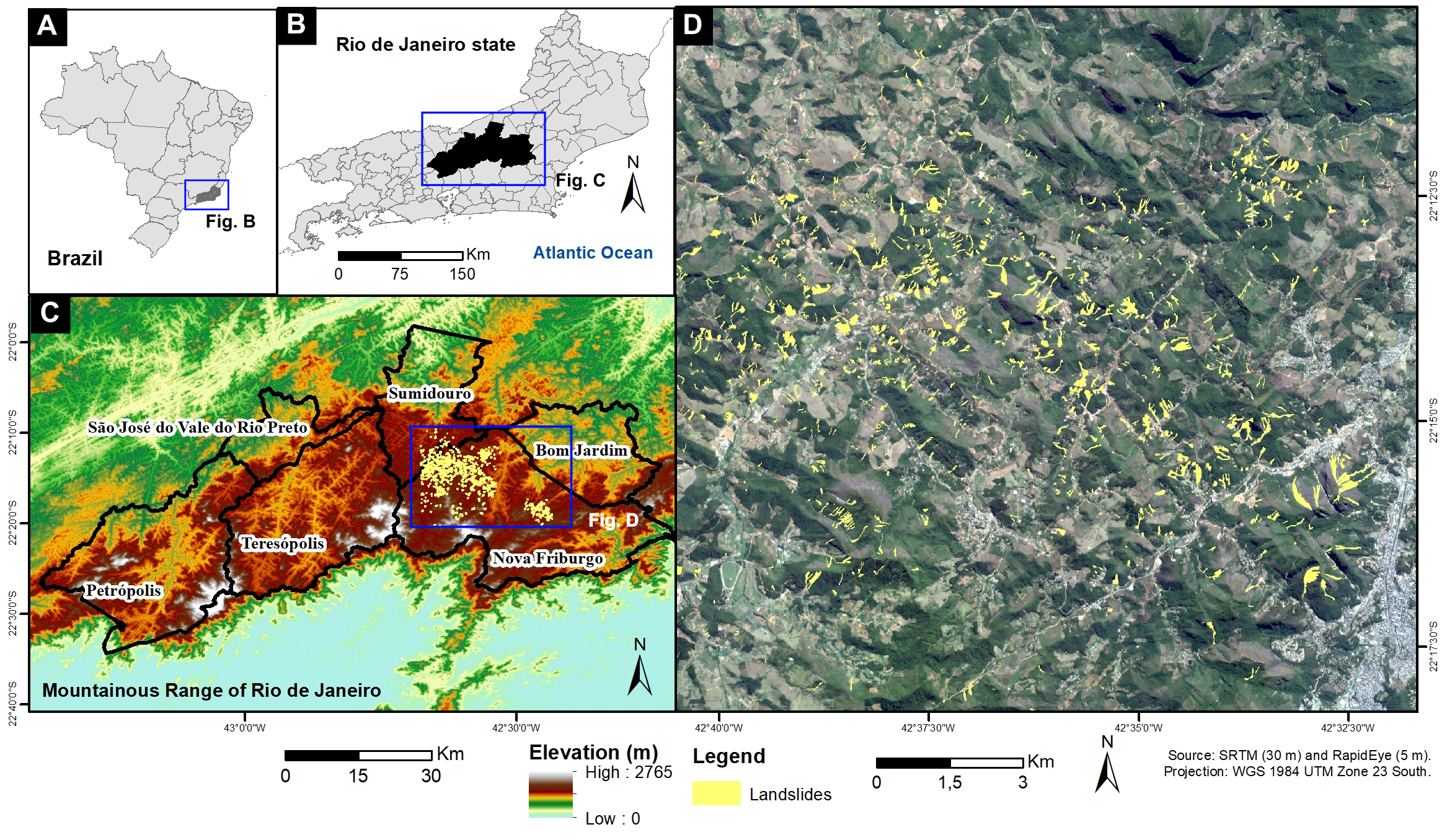}
\caption{A) Location of the study area in southeastern Brazil. B) Location of the Mountainous Range in the Rio de Janeiro State. C) Mountainous Range of Rio de Janeiro with the study area highlighted in blue. D) True color composition of the RapidEye image over Nova Friburgo, with the segmented landslides in yellow. The image was acquired on 2011-08-13.}
\label{fig:location}
\end{figure}

The mountainous region of Rio de Janeiro encompasses the municipalities of Nova Friburgo, Teresópolis, Petrópolis, Sumidouro, São José do Vale do Rio Preto and Bom Jardim (Fig.~\ref{fig:location}). The study area is in the municipality of Nova Friburgo, which was severely damaged by the disaster. 

Nova Friburgo is in the geomorphological unit of Serra dos Orgãos. The geological units have a WSW-ENE trend, and the elevation ranges between 1100 and 2000 meters above the mean sea level \citep{Dantas2001}. The geology consists mainly of igneous and metamorphic rocks such as granites, diorites, gabbros, and gneisses \citep{Tupinamba2012}. According to  Köppen's climate classification scheme \citep{Koppen1936}, the climate is subtropical highland (Cwb) with dry winter and mild summers. The annual mean precipitation is 1585.62~mm, with most of the rainfall in November, December, and January \citep{Sobral2018}.

\section{Methodology}
\label{sec:methods}

The spectral information from a RapidEye image and topographical information from a digital elevation model (DEM) derived from the ALOS's Phased Array L-band Synthetic Aperture Radar (PALSAR) were used to evaluate the performance of the U-net on landslide segmentation. The models were trained with images in three different window dimensions (32x32, 64x64, 128x128 pixels) that were sampled using random and regular grid methods. Random rotation, vertical, and horizontal flip were used for data augmentation. In total, 288 models were trained (Table~\ref{tbl:trained_models}). 

\begin{table}[!ht]
\centering
\caption{Number of trained models on each dataset.}
\label{tbl:trained_models}
\begin{tabular}{ll }
\hline
\textbf{\# of Models} & \textbf{Dataset}\\
\hline
72 & RapidEye \\

72 & RapidEye + Augmentation \\

72 & RapidEye + DEM\\

72 & RapidEye + DEM + Augmentation\\
\hline
\end{tabular}
\end{table}

The model's performance was evaluated in two test areas by using the mean intersect over union (mIoU), f1-score, precision, and recall metrics.  The proposed methodology involves the following steps: (1) data preprocessing, (2) model training (3) model evaluation.

\subsection{Data Preprocessing}
\subsubsection{RapidEye}
RapidEye consists of a constellation of five identical satellites with high-resolution sensors with a 6.5 meters nominal ground sampling distance at nadir. The orthorectified products are resampled and provided to users at a pixel size of 5 meters. The data are acquired with a temporal resolution of 5 days in five spectral bands: blue (440--510~nm), green (520--590~nm), red (630--685~nm), red-edge (690--730~nm), near-infrared (760--850~nm) \citep{rapideye2011satellite}. 

This work used the raw digital number (DN) of a 3A product (orthorectified, radiometric, and geometric corrections) with an area of 625~km\textsuperscript{2}. The image was acquired on 13 August 2011 and downloaded from the Planet Explorer website \citep{team2017planet}.

\subsubsection{ALOS/PALSAR}
The Phased Array type L-band Synthetic Aperture Radar (PALSAR) is one of the three observation sensors of the Advanced Land Observing Satellite (ALOS). PALSAR data is acquired at an off-nadir angle of 34.3 degrees in a Sun-synchronous Sub-recurrent Orbit (SSO) with a 46-day recurrent period. 

In this study, we used the radiometric terrain correction (RTC) product with 12.5 meters of spatial resolution obtained from the Alaska Satellite Facility (ASF) \citep{daac2015alos}. The image was acquired on 28 January 2011. The data were resampled to 5 meters pixel resolution with bilinear interpolation method, to match the spatial resolution of the RapidEye image, using GRASS GIS version 7.2.2 \citep{Neteler2012,GRASS_GIS_software}.

\subsubsection{Data Labeling}
The landslides were manually labeled in the RapidEye image using QGIS version 3.8 \citep{QGIS_software}. The segmented landslides were validated with Google Earth Pro version 7.3 \citep{googleearth} and by comparison with the landslide map produced by \citep{netto2013january}. In total, 1007 landslides were extracted from the scene, with area ranging from 200.32~m$^2$ to 78117.35~m$^2$ (638.51~m$^2$ average).  


\subsubsection{Test Areas}
\begin{figure}[!ht]
\centering
\includegraphics[width=0.9\textwidth, keepaspectratio]{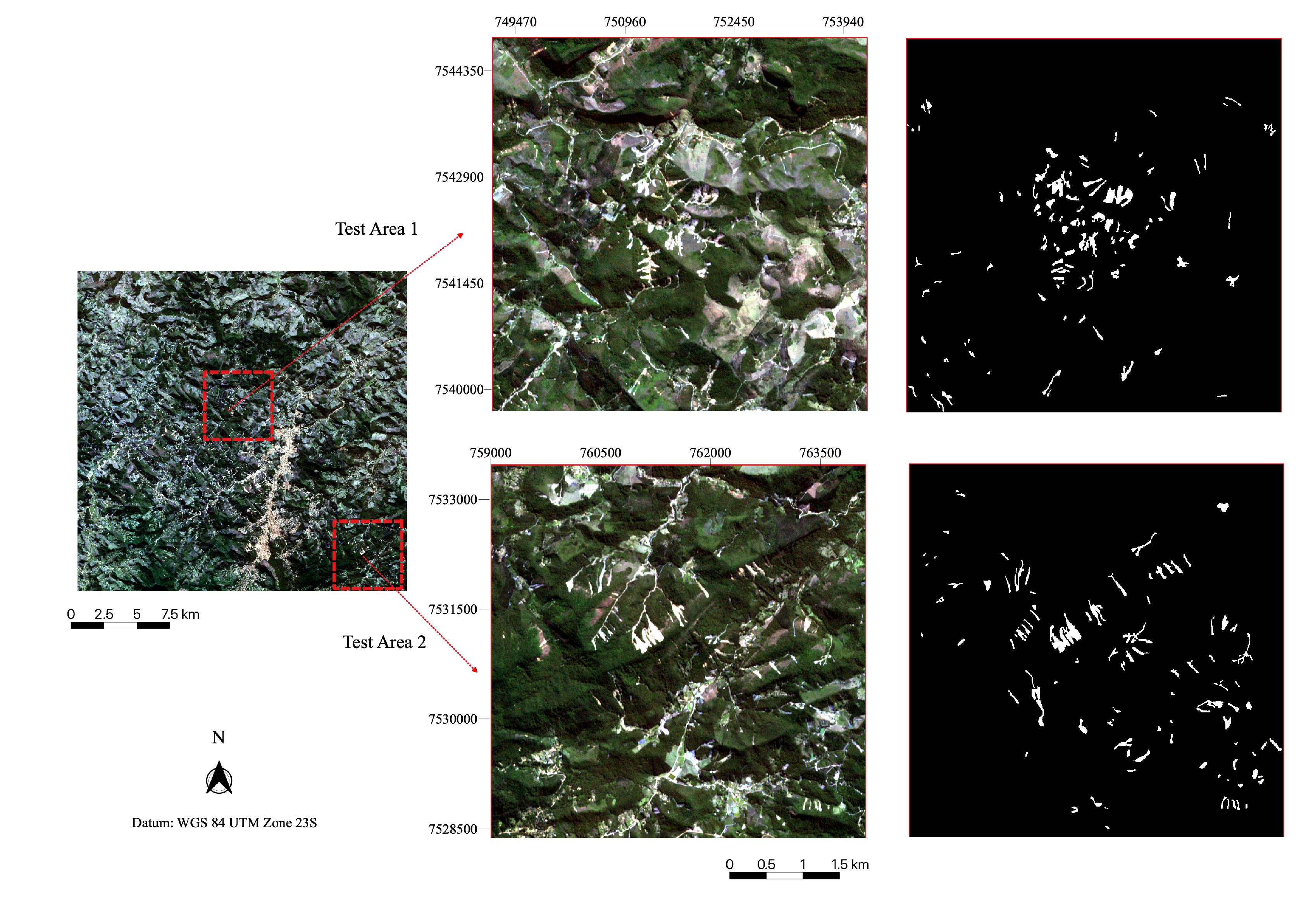}
\caption{Location of test areas and their binary masks. White pixels represent the manually segmented landslides and black pixels the background.}
\label{fig:train_areas}
\end{figure}

The model's accuracy was evaluated in two test areas with 1024x1024 pixels (Fig.~\ref{fig:train_areas}). In the first area, agriculture and grazing are predominant, while in the second, native vegetation and human settlements predominate. Ninety-six landslides were extracted from the first area and ninety-one from the second.

\subsubsection{Binary Mask}
The landslide scars polygons from the train and test areas were rasterized with Rasterio \citep{rasterio} and Numpy \citep{oliphant2006guide} Python libraries, to generate a binary mask with the same dimensions of the original scene. The pixels assigned with the value 1 (white pixels) correspond to the landslide scars class and the value 0 (black pixels) to the background class.

\subsubsection{Sampling Methods and Patch Sizes}
\begin{figure}[!ht]
\centering
\includegraphics[width=0.9\textwidth,keepaspectratio]{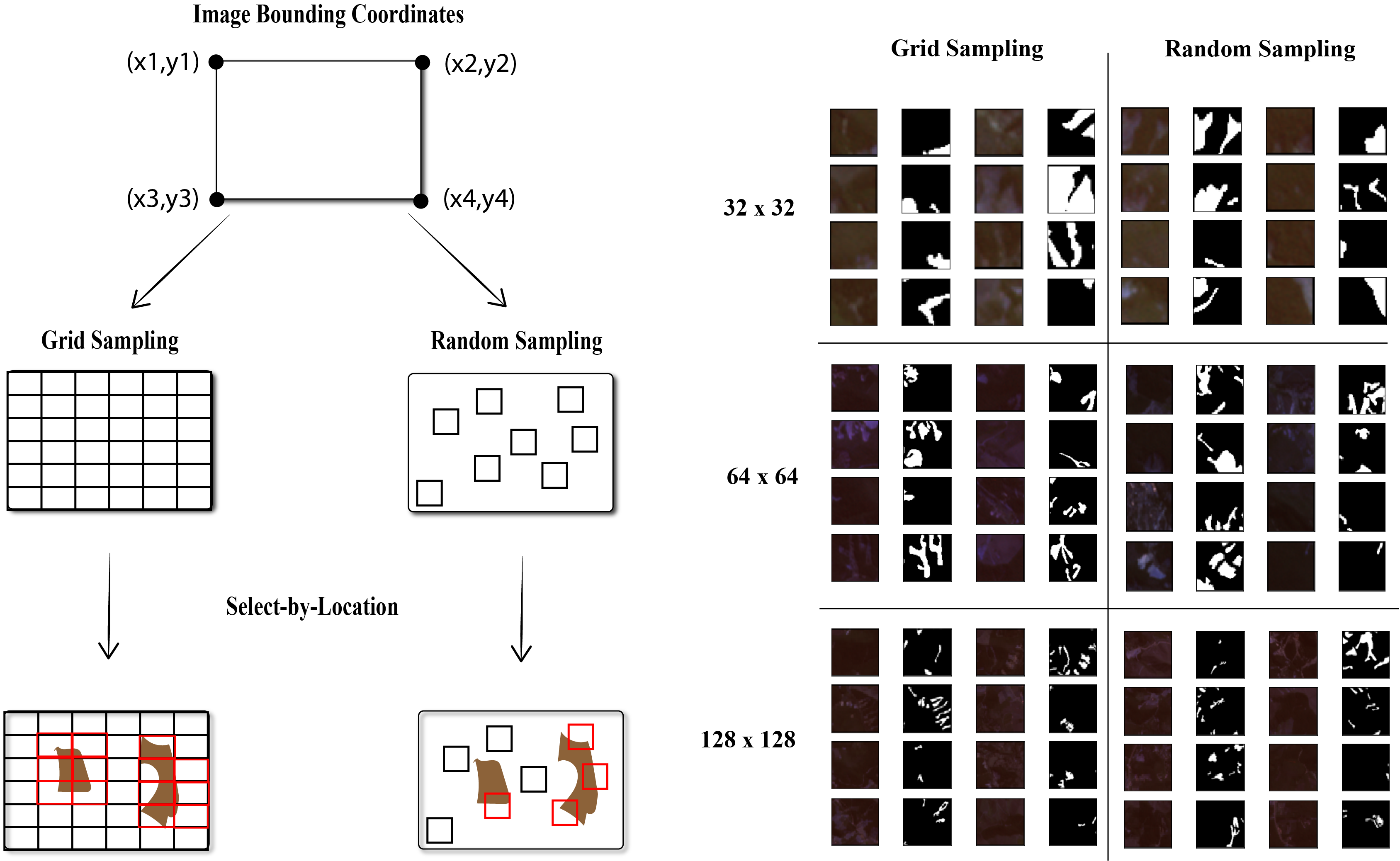}
\caption{Sampling methods (left) and results obtained for each patch size (right).}
\label{fig:test_sampling}
\end{figure}

The data was sampled by using random and regular grid methods in three different sizes: 32x32, 64x64, 128x128 pixels (Fig.~\ref{fig:test_sampling}).

The grid method used the bounding coordinates and the image resolution (5 meters) to generate a vector grid. The squares over the grid have an overlap of 20\%. The random sampling used the same procedure to generate 5000 sampling square polygons. A select-by-location operation was used to select only the polygons intersecting landslides. This ensures that all sampled images will have at least a small portion of a landslide scar, reducing class imbalance. The code used was adapted from the Keras-Spatial library \citep{KerasSpatial}.

\subsubsection{Data Augmentation}
The Albumentation library \citep{2018arXiv180906839B} was used to augment the data by using random rotations around 90$^{\circ}$, vertical, and horizontal flips. Table~\ref{tbl:patch_sizes} shows the sizes of all the datasets used to train the models.

\begin{table}[!ht]
\centering
\caption{Number of samples from all the datasets and patch sizes used to train the models.}
\label{tbl:patch_sizes}
\begin{tabular}{llll}
\hline
\textbf{Dataset} & \textbf{Size (pixels)} & \textbf{Regular Sampling} & \textbf{Random Sampling}\\
\hline
RapidEye                        &  32x32  &  3541  & 1740 \\
                                &  64x64  &  2264  & 2368 \\
RapidEye + DEM                  & 128x128 &  1565  & 3653 \\
\hline
RapidEye + Augmentation         &  32x32  &  9912  & 4872 \\
                                &  64x64  &  6336  & 6628 \\
RapidEye + DEM + Augmentation   & 128x128 &  4380  & 10228 \\
\hline
\end{tabular}
\end{table}

\subsection{Model Training}
\subsubsection{Model Architecture}
U-net \citep{ronneberger2015u} is a fully convolutional network developed for the segmentation of biomedical images. This type of architecture does not use fully connected layers in their structure; instead, they have an encoder-decoder architecture with just convolutional layers. The encoder path is responsible for classifying the pixels, but without taking the spatial location into account. The decoder path uses up-convolutions and concatenation to recover the spatial location of the classified pixels and return a mask with the same dimensions of the input image.

\begin{figure}[!ht]
\centering
\includegraphics[width=0.9\textwidth]{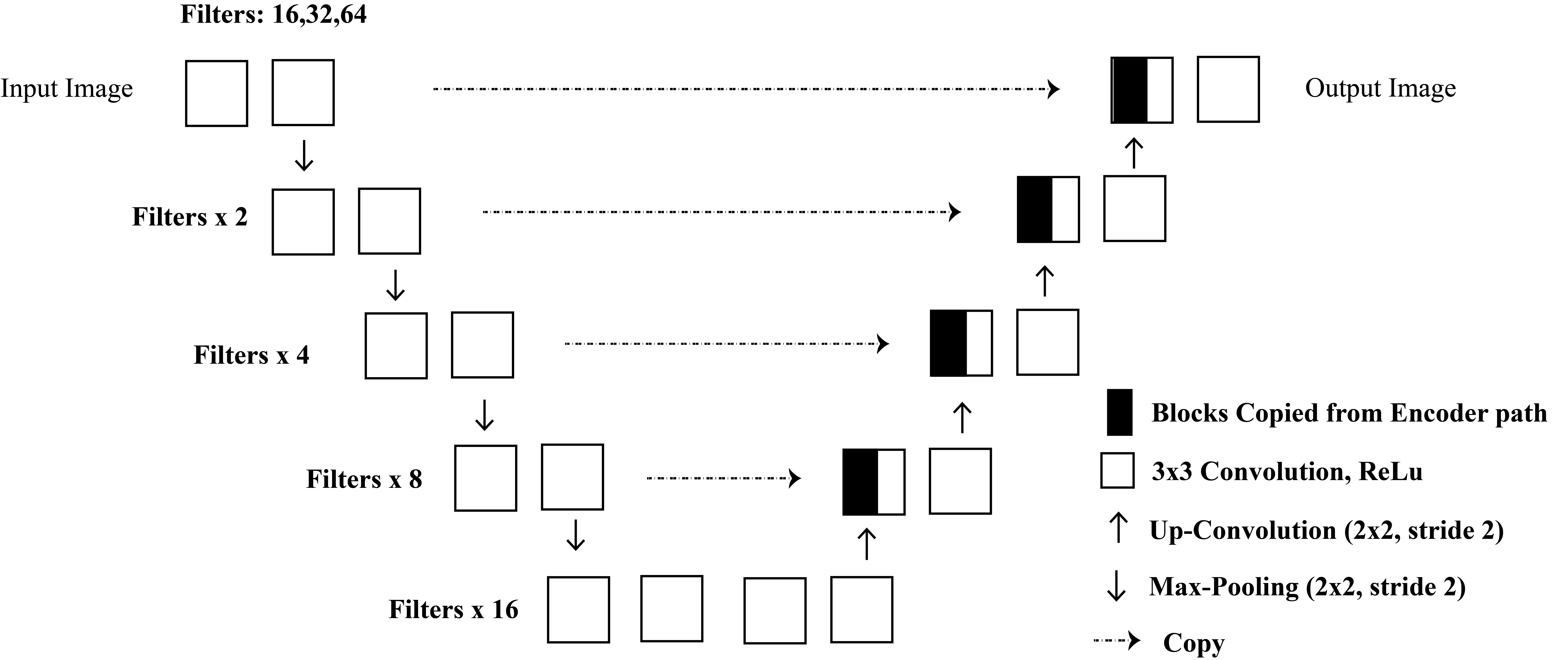}
\caption{U-net architecture. Filters increase by 2\textsuperscript{n} in each convolutional block.}
\label{fig:unet}
\end{figure}

In this study, we evaluated the U-net architecture (Fig.~\ref{fig:unet}) in three different values of initial filters: 16, 32, and 64 filters. The convolutional blocks on the encoder path have two 3x3 convolutional layers, activated with ReLu non-linear function, and followed by a max-pooling operation that reduces the spatial dimension by 2. In each convolutional block, the number of filters increases by  $2^{n}$, where $n$ is the block's position. On the decoder path,  2x2 up-sampling operations increase the data's spatial dimension to allow the concatenation of feature maps with the same dimension from the encoder path. Then, the concatenated data serve as input for two convolutional layers before another up-sampling operation. 

\subsubsection{Hyperparameters}

The models were trained for 200 epochs  with a fixed learning rate of 0.001. The initial tests also evaluated 0.01 and 0.0001 learning rate values, but the model's accuracy was lower than the models trained with a learning rate of 0.001. Binary cross-entropy and Adam were used as the loss and optimization function, respectively. The models were trained with four different batch sizes (16, 32, 64, 128 samples). The model's weights were just saved when the validation loss function decrease to reduce the overfitting. 

The models were trained on Google Colaboratory virtual environment \citep{googlecolab} with Keras \citep{chollet2015keras} and Tensorflow \citep{tensorflow2015-whitepaper} Python libraries. 30\% of each dataset was used as validation data. 

\subsection{Evaluation Metrics}
The model's performance was evaluated over two test areas by using f1-score, recall, precision, and mean intersect over union (mIoU) metrics. Those metrics are based on true positives (TP), false positives (FP), and false negatives (FN). TP are pixels correctly classified as landslides. FP represents the pixels incorrectly classified as landslides, and FN the pixels incorrectly classified as the background \citep{ghorbanzadeh2019evaluation,ghorbanzadeh2018dwelling,guirado2017deep}. The models that were trained with DEM as an additional channel were evaluated on test areas with an additional DEM channel. 

\subsubsection{Precision}
Precision (Eq.~\ref{eq:precision}) defines how accurate the model is by evaluating how much of the classified areas are landslides. The metric is useful for evaluating the cost of false positives.

\begin{equation}
 Precision  = \frac{True\:Positives}{True \: Positives + False\:Positives}
 \label{eq:precision}
\end{equation}

\subsubsection{Recall}
Recall (Eq.~\ref{eq:recall}) calculates how many of the actual positives are true positives. This metric is suitable to evaluate the cost associated with false negatives.

\begin{equation}
 Recal  = \frac{True\:Positives}{True \: Positives + False\:Negatives}
 \label{eq:recall}
\end{equation}

\subsubsection{F1-Score}
F1-score (Eq.~\ref{eq:f1}) combines precision and recall to measure if there is a balance between true positives and false negatives.

\begin{equation}
 F1-Score = 2*\frac{Precision\: * \:Recall }{Precision + Recall}
 \label{eq:f1}
\end{equation}

\subsubsection{Mean Intersect Over Union (mIoU)}
Mean intersect over union (Eq.~\ref{eq:miou}), also known as Jaccard Index, computes the overlapping of areas between the ground truth (A) and the model prediction (B) divided by the union of these areas. Then, the values are averaged for each class. A value of 1 (one) represents perfect overlapping, while 0 (zero) represents no overlap.  

\begin{equation}
 mIoU = \frac{A \cap B}{A \cup B} = \frac{True\: Positives}{True\: Positives + False\: Positives + False\: Negatives}
 \label{eq:miou}
\end{equation}

The result section shows for each dataset, sampling method, and patch size the models with the highest F1-Score and mIoU. The complete results are available in the Supplementary Material. The model generalization was evaluated by averaging the mIoU values from both test areas.

\section{Results and Discussion}
\begin{figure}[!ht]
\centering\includegraphics[width=0.9\textwidth]{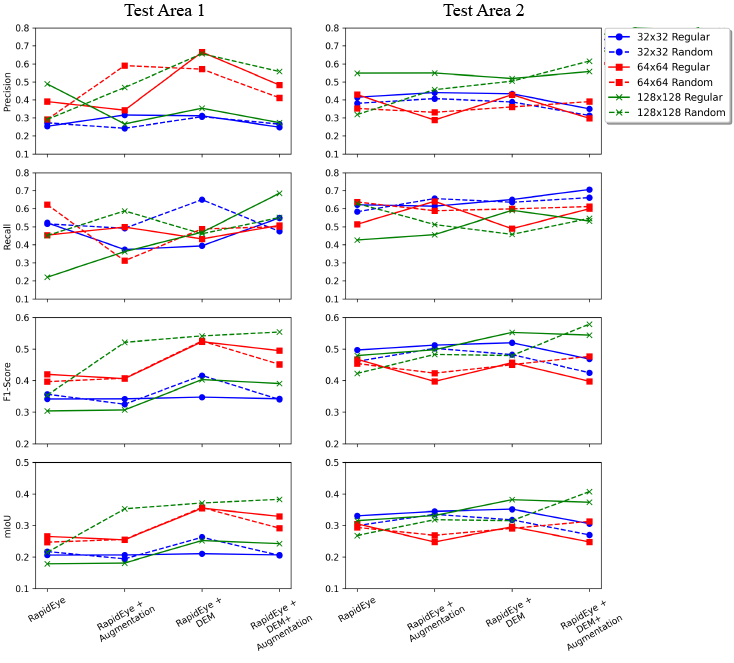}
  \caption{Precision, recall, f1-score and mIoU of the models with the highest f1-score and mIoU in test areas one (left) and two (right).}
  \label{fig:graph}
\end{figure}

The models were evaluated on two test areas with precision, recall, f1-score, and mIoU metrics. The results (Fig.~\ref{fig:graph}) shows that the models trained with the RapidEye+DEM and RapidEye+DEM+Augmentation datasets achieved the best results in all evaluated metrics in test area one. The models trained with 32x32 tiles had the lowest precision (0.24) over all the datasets, while models trained with regular 64x64 and random 128x128 tiles from the RapidEye+DEM dataset achieved 0.67 and 0.66 of precision. The recall was higher for the models trained with 128x128 regular tiles (0.68) and 32x32 random tiles (0.65). The model trained with 128x128 random tiles from the RapidEye+DEM+Augmentation dataset achieved the best f1-score (0.55) and mIoU (0.38).

In test area two, the dataset had a smaller influence over the results; however, models trained with RapidEye+DEM+  +Augmentation dataset also achieved the best results. The model trained with 128x128 random tiles achieved the highest precision (0.62), f1-score (0.58), and mIoU (0.41). Just recall was higher for the 32x32 regular tiles (0.70).  

\begin{figure}[!ht]
\centering
\includegraphics[width=0.9\textwidth]{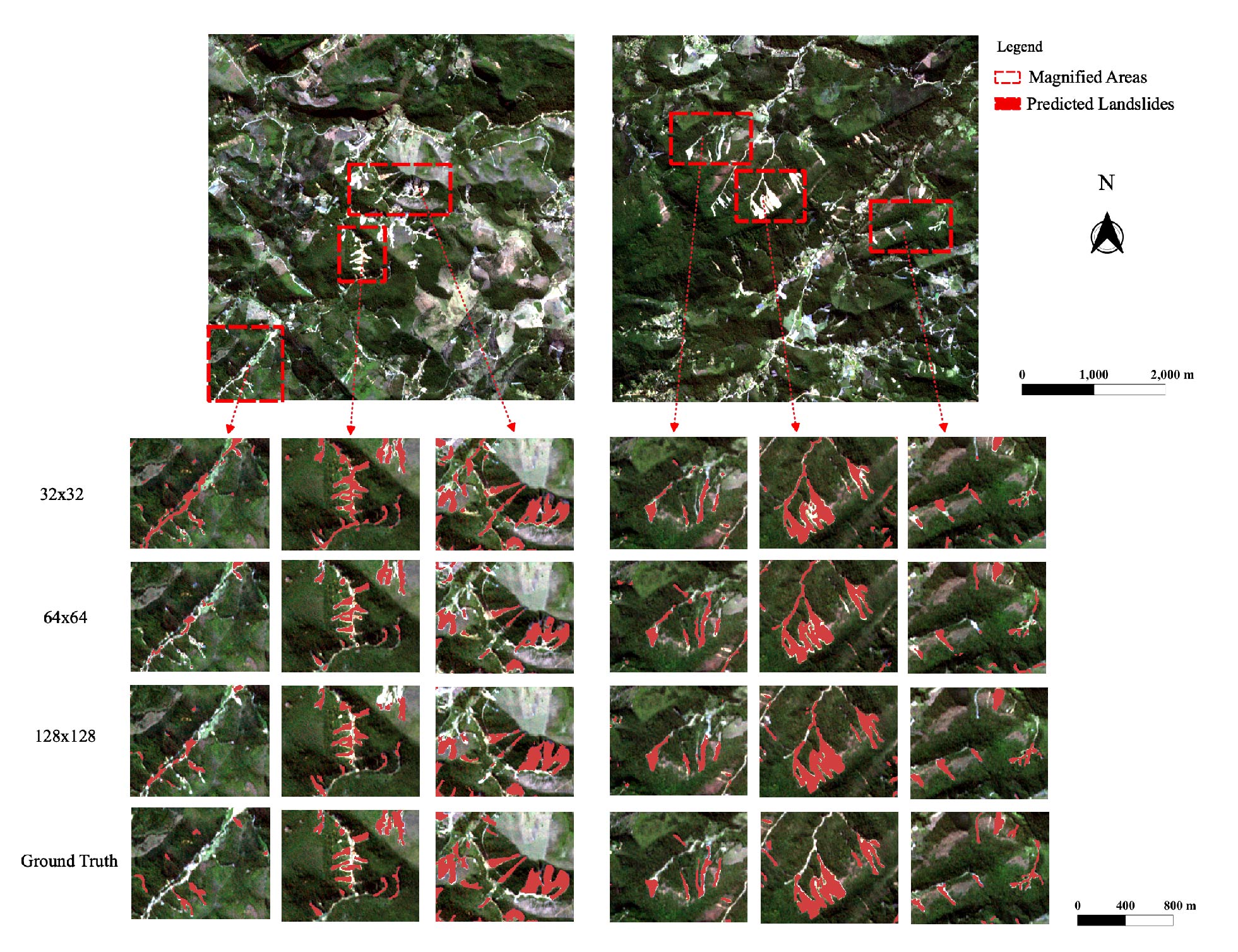}
\caption{Results of the models with the highest mIoU on each test area. In test area one (left), the image shows the models trained with 32x32 and 64x64 regular tiles from the RapidEye+DEM dataset, and 128x128 random tiles from the RapidEye+DEM+Augmentation dataset. In test are two (right), the image shows the models trained 32x32 and 64x64 regular tiles from RapidEye+DEM dataset, and the 128x128 model trained with random tiles from the RapidEye+DEM+Augmentation dataset.}
\label{fig:img_best_models}
\end{figure}

Precision evaluates the cost of false positives, while recall evaluates the cost of false negatives. At test area one (Fig.~\ref{fig:img_best_models}, left), the 32x32 models had lower precision meaning that they misclassify more background areas as landslides (false positives). Similar results occur at test area two (Fig.~\ref{fig:img_best_models}, right), but 64x64 models also had low precision values. Recall varied among the datasets; nevertheless, in both test areas, the models trained with 32x32 tiles achieved high results. Therefore, these models classified more landslide areas as landslides (true positive), reducing false negatives.

The F1-score value will always show a positive correlation with mIoU; however, mIoU tends to penalize incorrect classifications more quantitatively than f1-score. In both test areas, the models trained with random 128x128 tiles from the RapidEye+DEM+Augmentation dataset had the best f1-score and mIoU. In the test area one, this model achieved 0.55 of f1-score and 0.38 of mIoU. While in test area two, it achieved an f1-score of 0.58 and mIoU of 0.41. Comparing with the RapidEye dataset, where these models had the worst performance, the f1-score and mIoU increased 0.2 and 0.16 in test area one, and 0.16, 0.14 in test area two. 

When the test areas are evaluated individually, the sampling method seems to be less critical than the dataset to the overall accuracy of the models. However, by averaging the mIoU score of both test areas, it can be seen (Table~\ref{tbl:highest_avg_miou}) that random sampling outperformed the regular sampling. Thus, random sampling helped in increasing the generalization capacity of the models. Moreover, similar to what was observed in each test area individually, models trained with DEM and DEM+Augmentation had a better performance.

\begin{table}[!ht]
\centering
\caption{Results of the models with the highest average mIoU.}
\label{tbl:highest_avg_miou}
\begin{tabular}{*{14}{c}|}
\hline
\textbf{Sampling} & \textbf{Size} & \textbf{Test Area 1 - mIoU} & \textbf{Test Area 2 - mIoU}  & \textbf{Average mIoU} & \textbf{Dataset} \\
\hline
Random & 32 & 0.26 & 0.32 & 0.29 & RapidEye+DEM\\
Random & 64 & 0.29 & 0.31 & 0.30 & RapidEye+DEM+Augmentation\\
Random & 128 & 0.31 & 0.41 & 0.36 & RapidEye+DEM+Augmentation\\
\hline
\end{tabular}
\label{tab:label_test}
\end{table}

\begin{figure}[!ht]
\centering
\includegraphics[width=\textwidth]{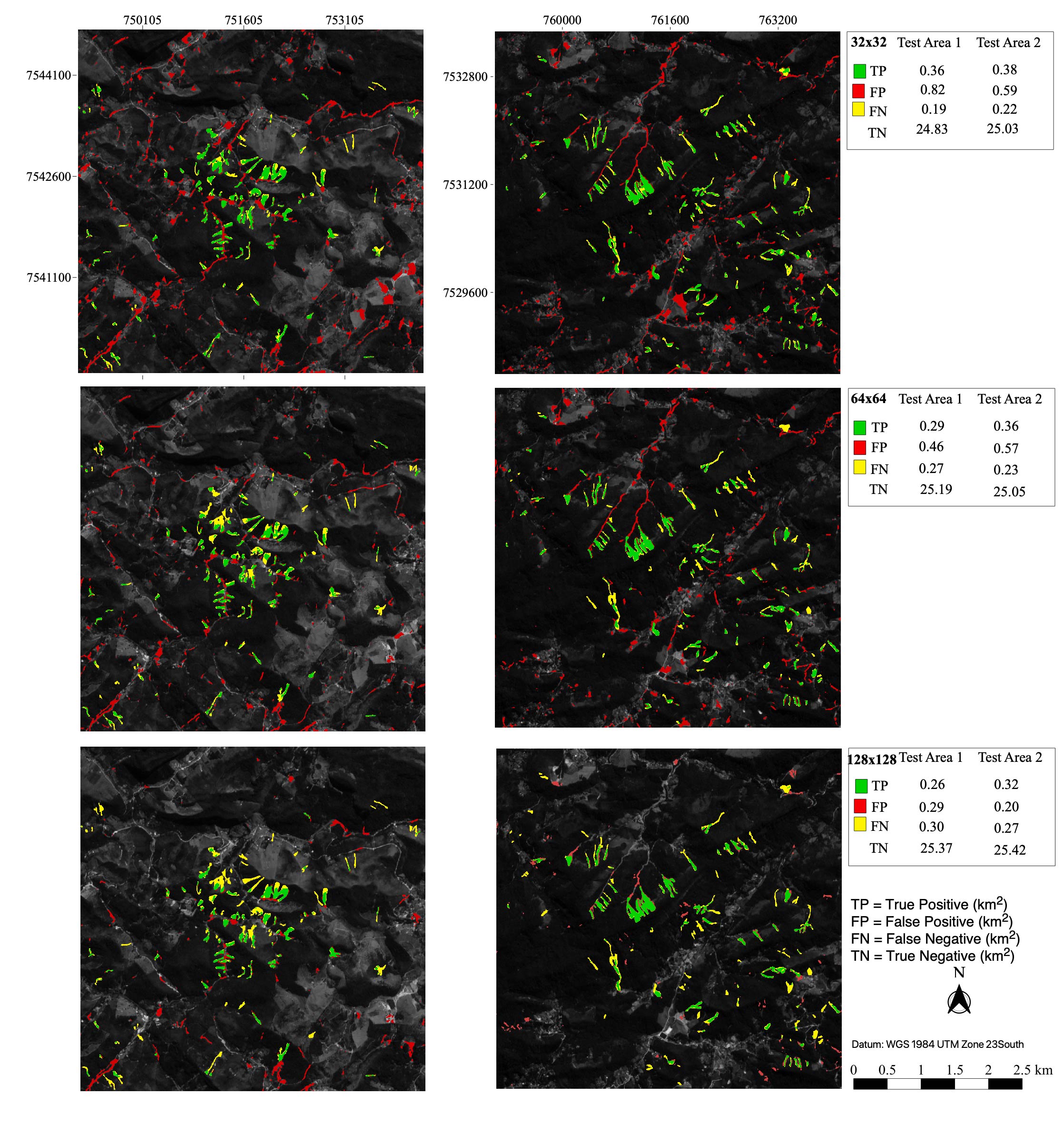}
\caption{Results of the models with the best generalization results (Table~\ref{tbl:highest_avg_miou}), for each patch size, on test area one (left) and two (right).}
\label{fig:img_best_general}
\end{figure}

The results (Fig.~\ref{fig:img_best_general}) of each model from table~\ref{tbl:highest_avg_miou}, shows that the 32x32 model predicted, in both test areas, 0.36 and 0.38~km$^2$ of true positives, 0.82 and 0.59 km\textsuperscript{2} of false positives, achieving the highest values. While the model trained with 128x128 tiles predicted the smallest true positive areas (0.26 and 0.32~km$^2$), false positive areas (0.29 and 0.20~km$^2$), and larger false negatives (0.30 and 0.27~km$^2$) and true negatives (25.37 and 25.42 ~m\textsuperscript{2}) areas. The model trained with 64x64 tiles achieved values in between those two models, with 0.29 and 0.36~km$^2$ of true positives; 0.46, 0.57~km$^2$ of false positives; 0.27 and 0.23~km$^2$ of false negatives, and true negative of 25.19 and 25.05~km$^2$.

The results of all evaluated models suggest that the models trained with smaller window sizes tend to understand the local context better. Thus, they classify more landslides correctly, achieving higher true positives and lower false negative values. However, as they are trained with small tiles, the scene's global context, which helps differentiate the background areas, is lost. As a result, they misclassify more background areas as landslides (false positives). On the other hand, models trained with the larger window sizes, in general, understand the global context better, reducing the false positive errors. Nevertheless, they classify a smaller number of pixels representing landslide scars correctly, increasing the number of false negatives.

Areas with similar spectral responses to landslides such as rivers with increased bedload, gravel roads, grazing, and agricultural areas are more common in area one than area two. Therefore, the models usually make more false positives errors in this area. In area two, the most common false positive mistakes were due to human settlements.

\section{Conclusions}

This study's main objective was to assess how the datasets, sampling methods, and patch sizes impact the overall performance of U-net on landslide segmentation. Our study suggests that the use of DEM and augmentation helped increase the overall accuracy of the models. Random sampling helped in increasing model generalization. Models trained with 32x32 patches classified more landslides areas correctly, thus, achieving higher true positive areas and lower false negatives. However, they also predict more false positive areas directly impacting precision, f1-score, and mIoU values. Contrarily, models trained with 128x128 tiles make less false positive errors and predict more areas correctly as background. Nevertheless, they also misclassify more landslide areas, increasing the number of false negatives and reducing the recall value. 64x64 tile models achieved results that lie in between the 32x32 and 128x128 models. 

In our study, the use of the digital elevation model as an additional channel helped improve the accuracy of the models. This results differs from the ones obtained by \citet{sameen2019landslide} and \citet{ghorbanzadeh2019evaluation}. However, since both authors used DEMs with higher spatial resolutions and the models were trained with different architectures and smaller tiles than the ones used in this research, more study is needed to address the most effective ways to use DEM with deep learning models. The 128x128 random model trained with the RapidEye+DEM+Augmentation dataset achieved the best performance in this research. The F1-score, 0.55 and 0.58, achieved in both test areas, is comparable to U-Net + ResNet34 evaluated by \citet{prakash2020mapping}, which achieved 0.56 of f1-score, and the PspNet tested by \citet{yu2020landslide} that achieved f1-score of  0.6. Future studies should explore multi-input models that can be trained with different input sizes; and evaluate different post-processing segmentation techniques to increase the quality of the results.

\section*{Computer Code Availability}
 All the codes used in this research are available on GitHub: \url{https://github.com/lpsmlgeobr/Landslide\_segmentation\_with\_unet}.

\section*{Data Availability}
The RapidEye image used in this study (Image ID:2328825) was acquired from Planet (\url{www.planet.com}) through Planet’s Education and Research Program. The ALOS PALSAR DEM (Tile 26708) is available from the Alaska Satellite Facility (ASF) Distributed Active Archive Center (DAAC -- \url{https://search.asf.alaska.edu/}).

\section*{Acknowledgements}

This study was supported by the Sao Paulo Research Foundation (FAPESP) grants \#2019/17555-1 and \#2016/06628-0 and by Brazil's National Council of Scientific and Technological Development, CNPq grants \#423481/2018-5 and \#304413/2018-6. This study was financed in part by CAPES Brasil - Finance Code 001.

\bibliographystyle{unsrtnat}

\bibliography{references}

\begin{thebibliography}{47}
\providecommand{\natexlab}[1]{#1}
\providecommand{\url}[1]{\texttt{#1}}
\expandafter\ifx\csname urlstyle\endcsname\relax
  \providecommand{\doi}[1]{doi: #1}\else
  \providecommand{\doi}{doi: \begingroup \urlstyle{rm}\Url}\fi

\bibitem[Kobiyama et~al.(2006)Kobiyama, Mendon{\c{c}}a, Moreno, Marcelino,
  Marcelino, Gon{\c{c}}alves, Brazetti, Goerl, Molleri, and
  Rudorff]{kobiyama2006prevenccao}
Masato Kobiyama, Magaly Mendon{\c{c}}a, Davis~Anderson Moreno, IPVO Marcelino,
  Emerson~V Marcelino, Edison~F Gon{\c{c}}alves, Leticia~LP Brazetti, Roberto~F
  Goerl, Gustavo~SF Molleri, and F~de~M Rudorff.
\newblock \emph{Preven{\c{c}}{\~a}o de desastres naturais: conceitos
  b{\'a}sicos}.
\newblock Organic Trading Curitiba, 2006.

\bibitem[Hong et~al.(2017)Hong, Chen, Xu, Youssef, Pradhan, and
  Tien~Bui]{hong2017rainfall}
Haoyuan Hong, Wei Chen, Chong Xu, Ahmed~M Youssef, Biswajeet Pradhan, and Dieu
  Tien~Bui.
\newblock Rainfall-induced landslide susceptibility assessment at the chongren
  area (china) using frequency ratio, certainty factor, and index of entropy.
\newblock \emph{Geocarto international}, 32\penalty0 (2):\penalty0 139--154,
  2017.

\bibitem[Alexander(2008)]{alexander2008brief}
David~E Alexander.
\newblock A brief survey of gis in mass-movement studies, with reflections on
  theory and methods.
\newblock \emph{Geomorphology}, 94\penalty0 (3-4):\penalty0 261--267, 2008.

\bibitem[Zhong et~al.(2019)Zhong, Liu, Gao, Chen, Li, Hou, Nuremanguli, and
  Ma]{zhong2019landslide}
Cheng Zhong, Yue Liu, Peng Gao, Wenlong Chen, Hui Li, Yong Hou, Tuohuti
  Nuremanguli, and Haijian Ma.
\newblock Landslide mapping with remote sensing: challenges and opportunities.
\newblock \emph{International Journal of Remote Sensing}, pages 1--27, 2019.

\bibitem[UNISDR(2015)]{unisdr2015sendai}
UNO UNISDR.
\newblock Sendai framework for disaster risk reduction 2015--2030.
\newblock In \emph{Proceedings of the 3rd United Nations World Conference on
  DRR, Sendai, Japan}, pages 14--18, 2015.

\bibitem[Mondini et~al.(2019)Mondini, Santangelo, Rocchetti, Rossetto, Manconi,
  and Monserrat]{mondini2019sentinel}
Alessandro~C Mondini, Michele Santangelo, Margherita Rocchetti, Enrica
  Rossetto, Andrea Manconi, and Oriol Monserrat.
\newblock Sentinel-1 sar amplitude imagery for rapid landslide detection.
\newblock \emph{Remote Sensing}, 11\penalty0 (7):\penalty0 760, 2019.

\bibitem[Guzzetti et~al.(2012)Guzzetti, Mondini, Cardinali, Fiorucci,
  Santangelo, and Chang]{guzzetti2012landslide}
Fausto Guzzetti, Alessandro~Cesare Mondini, Mauro Cardinali, Federica Fiorucci,
  Michele Santangelo, and Kang-Tsung Chang.
\newblock Landslide inventory maps: New tools for an old problem.
\newblock \emph{Earth-Science Reviews}, 112\penalty0 (1-2):\penalty0 42--66,
  2012.

\bibitem[Van~Westen et~al.(2008)Van~Westen, Castellanos, and
  Kuriakose]{VanWesten2008}
Cees~J Van~Westen, Enrique Castellanos, and Sekhar~L Kuriakose.
\newblock Spatial data for landslide susceptibility, hazard, and vulnerability
  assessment: an overview.
\newblock \emph{Engineering geology}, 102\penalty0 (3-4):\penalty0 112--131,
  2008.

\bibitem[Zhong et~al.(2020)Zhong, Liu, Gao, Chen, Li, Hou, Nuremanguli, and
  Ma]{zhong2020landslide}
Cheng Zhong, Yue Liu, Peng Gao, Wenlong Chen, Hui Li, Yong Hou, Tuohuti
  Nuremanguli, and Haijian Ma.
\newblock Landslide mapping with remote sensing: challenges and opportunities.
\newblock \emph{International Journal of Remote Sensing}, 41\penalty0
  (4):\penalty0 1555--1581, 2020.

\bibitem[Xu(2015)]{xu2015preparation}
Chong Xu.
\newblock Preparation of earthquake-triggered landslide inventory maps using
  remote sensing and gis technologies: Principles and case studies.
\newblock \emph{Geoscience Frontiers}, 6\penalty0 (6):\penalty0 825--836, 2015.

\bibitem[Yu et~al.(2020)Yu, Chen, and Xu]{yu2020landslide}
Bo~Yu, Fang Chen, and Chong Xu.
\newblock Landslide detection based on contour-based deep learning framework in
  case of national scale of nepal in 2015.
\newblock \emph{Computers \& Geosciences}, 135:\penalty0 104388, 2020.

\bibitem[Blaschke(2010)]{Blaschke2010}
Thomas Blaschke.
\newblock Object based image analysis for remote sensing.
\newblock \emph{ISPRS journal of photogrammetry and remote sensing},
  65\penalty0 (1):\penalty0 2--16, 2010.

\bibitem[Stumpf and Kerle(2011)]{stumpf2011object}
Andr{\'e} Stumpf and Norman Kerle.
\newblock Object-oriented mapping of landslides using random forests.
\newblock \emph{Remote sensing of environment}, 115\penalty0 (10):\penalty0
  2564--2577, 2011.

\bibitem[Blaschke et~al.(2014)Blaschke, Hay, Kelly, Lang, Hofmann, Addink,
  Feitosa, Van~der Meer, Van~der Werff, Van~Coillie,
  et~al.]{blaschke2014geographic}
Thomas Blaschke, Geoffrey~J Hay, Maggi Kelly, Stefan Lang, Peter Hofmann,
  Elisabeth Addink, Raul~Queiroz Feitosa, Freek Van~der Meer, Harald Van~der
  Werff, Frieke Van~Coillie, et~al.
\newblock Geographic object-based image analysis--towards a new paradigm.
\newblock \emph{ISPRS journal of photogrammetry and remote sensing},
  87:\penalty0 180--191, 2014.

\bibitem[Prakash et~al.(2020)Prakash, Manconi, and Loew]{prakash2020mapping}
Nikhil Prakash, Andrea Manconi, and Simon Loew.
\newblock Mapping landslides on eo data: Performance of deep learning models
  vs. traditional machine learning models.
\newblock \emph{Remote Sensing}, 12\penalty0 (3):\penalty0 346, 2020.

\bibitem[Ghorbanzadeh et~al.(2019)Ghorbanzadeh, Blaschke, Gholamnia, Meena,
  Tiede, and Aryal]{ghorbanzadeh2019evaluation}
Omid Ghorbanzadeh, Thomas Blaschke, Khalil Gholamnia, Sansar~Raj Meena, Dirk
  Tiede, and Jagannath Aryal.
\newblock Evaluation of different machine learning methods and deep-learning
  convolutional neural networks for landslide detection.
\newblock \emph{Remote Sensing}, 11\penalty0 (2):\penalty0 196, 2019.

\bibitem[Peng et~al.(2019)Peng, Zhang, and Guan]{peng2019end}
Daifeng Peng, Yongjun Zhang, and Haiyan Guan.
\newblock End-to-end change detection for high resolution satellite images
  using improved unet++.
\newblock \emph{Remote Sensing}, 11\penalty0 (11):\penalty0 1382, 2019.

\bibitem[Zhu et~al.(2017)Zhu, Tuia, Mou, Xia, Zhang, Xu, and
  Fraundorfer]{zhu2017deep}
Xiao~Xiang Zhu, Devis Tuia, Lichao Mou, Gui-Song Xia, Liangpei Zhang, Feng Xu,
  and Friedrich Fraundorfer.
\newblock Deep learning in remote sensing: A comprehensive review and list of
  resources.
\newblock \emph{IEEE Geoscience and Remote Sensing Magazine}, 5\penalty0
  (4):\penalty0 8--36, 2017.

\bibitem[Long et~al.(2015)Long, Shelhamer, and Darrell]{long2015fully}
Jonathan Long, Evan Shelhamer, and Trevor Darrell.
\newblock Fully convolutional networks for semantic segmentation.
\newblock In \emph{Proceedings of the IEEE conference on computer vision and
  pattern recognition}, pages 3431--3440, 2015.

\bibitem[Radovic et~al.(2017)Radovic, Adarkwa, and Wang]{radovic2017object}
Matija Radovic, Offei Adarkwa, and Qiaosong Wang.
\newblock Object recognition in aerial images using convolutional neural
  networks.
\newblock \emph{Journal of Imaging}, 3\penalty0 (2):\penalty0 21, 2017.

\bibitem[Ding et~al.(2016)Ding, Zhang, Zhou, and Dai]{ding2016automatic}
Anzi Ding, Qingyong Zhang, Xinmin Zhou, and Bicheng Dai.
\newblock Automatic recognition of landslide based on cnn and texture change
  detection.
\newblock In \emph{2016 31st Youth Academic Annual Conference of Chinese
  Association of Automation (YAC)}, pages 444--448. IEEE, 2016.

\bibitem[Chen et~al.(2018)Chen, Zhang, Ouyang, Zhang, and
  Ma]{chen2018automated}
Zhong Chen, Yifei Zhang, Chao Ouyang, Feng Zhang, and Jie Ma.
\newblock Automated landslides detection for mountain cities using
  multi-temporal remote sensing imagery.
\newblock \emph{Sensors}, 18\penalty0 (3):\penalty0 821, 2018.

\bibitem[Sameen and Pradhan(2019)]{sameen2019landslide}
Maher~Ibrahim Sameen and Biswajeet Pradhan.
\newblock Landslide detection using residual networks and the fusion of
  spectral and topographic information.
\newblock \emph{IEEE Access}, 7:\penalty0 114363--114373, 2019.

\bibitem[Pradhan et~al.(2017)Pradhan, Seeni, and
  Nampak]{pradhan2017integration}
Biswajeet Pradhan, Maher~Ibrahim Seeni, and Haleh Nampak.
\newblock Integration of lidar and quickbird data for automatic landslide
  detection using object-based analysis and random forests.
\newblock In \emph{Laser Scanning Applications in Landslide Assessment}, pages
  69--81. Springer, 2017.

\bibitem[Avelar et~al.(2013)Avelar, Netto, Lacerda, Becker, and
  Mendon{\c{c}}a]{avelar2013mechanisms}
Andr{\'e}~S Avelar, Ana L~Coelho Netto, Willy~A Lacerda, Leonardo~B Becker, and
  Marcos~B Mendon{\c{c}}a.
\newblock Mechanisms of the recent catastrophic landslides in the mountainous
  range of rio de janeiro, brazil.
\newblock In \emph{Landslide science and practice}, pages 265--270. Springer,
  2013.

\bibitem[Dantas(2001)]{Dantas2001}
M.~E. Dantas.
\newblock Geomorfologia do estado do rio de janeiro.
\newblock \emph{CPRM. Estudo geoambiental do Estado do Rio de Janeiro.
  Bras{\'\i}lia}, 2001.

\bibitem[Tupinambá et~al.(2012)Tupinambá, Heilbron, Duarte, de~Almeida,
  Valladares, Pacheco, dos Santos~Salomão, Conceição, da~Silva, de~Almeida,
  Ferrassoli, de~Cássia O.~da Costa, Tupinambá, Rocha, Benac, da~Silva,
  Guimarães, (DRM-RJ), (DRM-RJ), de~Lima~da Silva, Palermo, and
  Pereira]{Tupinamba2012}
Miguel Tupinambá, Monica Heilbron, Beatriz~Paschoal Duarte, Julio Cesar~Horta
  de~Almeida, Claudia~Sayão Valladares, Bruno~Trota Pacheco, Marcelo dos
  Santos~Salomão, Flávio~Ribeiro Conceição, Luiz Guilherme~Eirado da~Silva,
  Clayton~Guia de~Almeida, Marcelo~Ambrósio Ferrassoli, Mariana de~Cássia
  O.~da Costa, Luciana~Rocha Tupinambá, David~Silva Rocha, Pedro~Monteiro
  Benac, Hugo Mathias O.~Carvalho da~Silva, Paulo~Vicente Guimarães, (DRM-RJ),
  (DRM-RJ), Felipe de~Lima~da Silva, Nely Palermo, and Ronaldo~Mello Pereira.
\newblock Mapa geológico folha nova friburgo sf-23-z-b-ii.
\newblock Technical report, CPRM - Serviço Geológico do Brasil, 2012.

\bibitem[Köppen(1936)]{Koppen1936}
W.~Köppen.
\newblock \emph{Das geographische System der Klimate}, volume~1, chapter Das
  geographische System der Klimate, pages 1--44.
\newblock Gerbrüder Bornträger, 1936.

\bibitem[Sobral et~al.(2018)Sobral, Oliveira-J{\'u}nior, Gois,
  de~Bodas~Terassi, and Muniz-J{\'u}nior]{Sobral2018}
Bruno~Serafini Sobral, Jos{\'e}~Francisco Oliveira-J{\'u}nior, Givanildo Gois,
  Paulo~Miguel de~Bodas~Terassi, and Jo{\~a}o Gualberto~Rodrigues
  Muniz-J{\'u}nior.
\newblock Variabilidade espa{\c{c}}o-temporal e interanual da chuva no estado
  do rio de janeiro.
\newblock \emph{Revista Brasileira de Climatologia}, 22, 2018.

\bibitem[RapidEye(2011)]{rapideye2011satellite}
AG~RapidEye.
\newblock Satellite imagery product specifications.
\newblock \emph{Satellite imagery product specifications: Version}, 2011.

\bibitem[{Planet Team}(2017)]{team2017planet}
{Planet Team}.
\newblock Planet application program interface: In space for life on earth. san
  francisco, ca, 2017.
\newblock Available online: \url{https://api.planet.com.} Last accessed on
  2020-05-1.

\bibitem[DAAC(2015)]{daac2015alos}
ASF DAAC.
\newblock Alos palsar\_radiometric\_terrain\_corrected\_high\_res.
\newblock \emph{JAXA/METI}, 11, 2015.
\newblock \url{https://search.asf.alaska.edu/\#/}, Last accessed on 2019-12-05.

\bibitem[Neteler et~al.(2012)Neteler, Bowman, Landa, and Metz]{Neteler2012}
Markus Neteler, M.~Hamish Bowman, Martin Landa, and Markus Metz.
\newblock {GRASS GIS: A multi-purpose open source GIS}.
\newblock \emph{Environmental Modelling \& Software}, 31:\penalty0 124--130,
  2012.
\newblock ISSN 1364-8152.
\newblock \doi{10.1016/j.envsoft.2011.11.014}.

\bibitem[{GRASS Development Team}(2017)]{GRASS_GIS_software}
{GRASS Development Team}.
\newblock \emph{Geographic Resources Analysis Support System (GRASS GIS)
  Software, Version 7.2}.
\newblock Open Source Geospatial Foundation, 2017.
\newblock URL \url{http://grass.osgeo.org}.

\bibitem[{QGIS Development Team}(2009)]{QGIS_software}
{QGIS Development Team}.
\newblock \emph{QGIS Geographic Information System}.
\newblock Open Source Geospatial Foundation, 2009.
\newblock URL \url{http://qgis.osgeo.org}.

\bibitem[Google(2019)]{googleearth}
Google.
\newblock Google earth pro version 7.3, 2019.
\newblock \url{https://www.google.com/earth/versions/\#download-pro}, Last
  accessed on 2020-04-20.

\bibitem[Netto et~al.(2013)Netto, Sato, de~Souza~Avelar, Vianna, Ara{\'u}jo,
  Ferreira, Lima, Silva, and Silva]{netto2013january}
Ana Luiza~Coelho Netto, Anderson~Mululo Sato, Andr{\'e} de~Souza~Avelar,
  L{\'\i}lian Gabriela~G Vianna, Ingrid~S Ara{\'u}jo, David~LC Ferreira,
  Pedro~H Lima, Ana Paula~A Silva, and Roberta~P Silva.
\newblock January 2011: the extreme landslide disaster in brazil.
\newblock In \emph{Landslide science and practice}, pages 377--384. Springer,
  2013.

\bibitem[Gillies et~al.(2013--)]{rasterio}
Sean Gillies et~al.
\newblock Rasterio: geospatial raster i/o for {Python} programmers, 2013--.
\newblock URL \url{https://github.com/mapbox/rasterio}.

\bibitem[Oliphant(2006)]{oliphant2006guide}
Travis~E Oliphant.
\newblock \emph{A guide to NumPy}, volume~1.
\newblock Trelgol Publishing USA, 2006.

\bibitem[Terstriep(2019)]{KerasSpatial}
Jeff Terstriep.
\newblock Keras spatial, 2019.
\newblock \url{https://pypi.org/project/keras-spatial/}, Last accessed on
  2020-02-10.

\bibitem[Buslaev et~al.(2018)Buslaev, Parinov, Khvedchenya, Iglovikov, and
  Kalinin]{2018arXiv180906839B}
A.~Buslaev, A.~Parinov, E.~Khvedchenya, V.~I. Iglovikov, and A.~A. Kalinin.
\newblock {Albumentations: fast and flexible image augmentations}.
\newblock \emph{ArXiv e-prints}, 2018.

\bibitem[Ronneberger et~al.(2015)Ronneberger, Fischer, and
  Brox]{ronneberger2015u}
Olaf Ronneberger, Philipp Fischer, and Thomas Brox.
\newblock U-net: Convolutional networks for biomedical image segmentation.
\newblock In \emph{International Conference on Medical image computing and
  computer-assisted intervention}, pages 234--241. Springer, 2015.

\bibitem[Google(2018)]{googlecolab}
Google.
\newblock Google colaboratory, 2018.
\newblock \url{https://colab.research.google.com/}, Last accessed on
  2020-05-12.

\bibitem[Chollet et~al.(2015)]{chollet2015keras}
Fran\c{c}ois Chollet et~al.
\newblock Keras.
\newblock \url{https://github.com/fchollet/keras}, 2015.

\bibitem[Abadi et~al.(2015)Abadi, Agarwal, Barham, Brevdo, Chen, Citro,
  Corrado, Davis, Dean, Devin, Ghemawat, Goodfellow, Harp, Irving, Isard, Jia,
  Jozefowicz, Kaiser, Kudlur, Levenberg, Man\'{e}, Monga, Moore, Murray, Olah,
  Schuster, Shlens, Steiner, Sutskever, Talwar, Tucker, Vanhoucke, Vasudevan,
  Vi\'{e}gas, Vinyals, Warden, Wattenberg, Wicke, Yu, and
  Zheng]{tensorflow2015-whitepaper}
Mart\'{\i}n Abadi, Ashish Agarwal, Paul Barham, Eugene Brevdo, Zhifeng Chen,
  Craig Citro, Greg~S. Corrado, Andy Davis, Jeffrey Dean, Matthieu Devin,
  Sanjay Ghemawat, Ian Goodfellow, Andrew Harp, Geoffrey Irving, Michael Isard,
  Yangqing Jia, Rafal Jozefowicz, Lukasz Kaiser, Manjunath Kudlur, Josh
  Levenberg, Dandelion Man\'{e}, Rajat Monga, Sherry Moore, Derek Murray, Chris
  Olah, Mike Schuster, Jonathon Shlens, Benoit Steiner, Ilya Sutskever, Kunal
  Talwar, Paul Tucker, Vincent Vanhoucke, Vijay Vasudevan, Fernanda Vi\'{e}gas,
  Oriol Vinyals, Pete Warden, Martin Wattenberg, Martin Wicke, Yuan Yu, and
  Xiaoqiang Zheng.
\newblock {TensorFlow}: Large-scale machine learning on heterogeneous systems,
  2015.
\newblock URL \url{https://www.tensorflow.org/}.
\newblock Software available from tensorflow.org.

\bibitem[Ghorbanzadeh et~al.(2018)Ghorbanzadeh, Tiede, Dabiri, Sudmanns, and
  Lang]{ghorbanzadeh2018dwelling}
Omid Ghorbanzadeh, Dirk Tiede, Zahra Dabiri, Martin Sudmanns, and Stefan Lang.
\newblock Dwelling extraction in refugee camps using cnn--first experiences and
  lessons learnt.
\newblock \emph{International Archives of the Photogrammetry, Remote Sensing \&
  Spatial Information Sciences}, 2018.

\bibitem[Guirado et~al.(2017)Guirado, Tabik, Alcaraz-Segura, Cabello, and
  Herrera]{guirado2017deep}
Emilio Guirado, Siham Tabik, Domingo Alcaraz-Segura, Javier Cabello, and
  Francisco Herrera.
\newblock Deep-learning convolutional neural networks for scattered shrub
  detection with google earth imagery.
\newblock \emph{arXiv preprint arXiv:1706.00917}, 2017.

\end{thebibliography}

\end{document}